\documentclass[aps,twocolumn,showpacs,floatfix,superscriptaddress]{revtex4}
\usepackage{graphicx}
\usepackage{amsmath}
\usepackage{amssymb}

\begin{document}
\def\be{\begin{equation}}
\def\ee{\end{equation}}
\def\bea{\begin{eqnarray}}
\def\eea{\end{eqnarray}}
\def\bma{\begin{mathletters}}
\def\ema{\end{mathletters}}
\newcommand{\one}{\mbox{$1 \hspace{-1.0mm}  {\bf l}$}}
\newcommand{\eins}{\mbox{$1 \hspace{-1.0mm}  {\bf l}$}}
\def\C{\hbox{$\mit I$\kern-.7em$\mit C$}}
\newcommand{\tr}{{\rm tr}}
\newcommand{\half}{\mbox{$\textstyle \frac{1}{2}$}}
\newcommand{\shalf}{\mbox{$\textstyle \frac{1}{\sqrt{2}}$}}
\newcommand{\ket}[1]{ | \, #1  \rangle}
\newcommand{\bra}[1]{ \langle #1 \,  |}
\newcommand{\proj}[1]{\ket{#1}\bra{#1}}
\newcommand{\kb}[2]{\ket{#1}\bra{#2}}
\newcommand{\bk}[2]{\langle \, #1 | \, #2 \rangle}
\def\II{I(\{p_k\},\{\rho_k\})}
\def\ss{{\cal K}}
\tolerance = 10000




\title{Exploiting Quantum Parallelism \\To Simulate
Quantum Random Many-Body Systems}

\author{B. Paredes}
\affiliation{Max--Planck Institut f\"{u}r Quantenoptik,
Hans-KopfermannStr. 1, Garching, D-85748 Germany}
\author{F. Verstraete}
\affiliation{Institute for Quantum Information, Caltech, Pasadena
91125, CA}
\author{J. I. Cirac}
\affiliation{Max--Planck Institut f\"{u}r Quantenoptik,
Hans-KopfermannStr. 1, Garching, D-85748 Germany}

\begin{abstract}
We present an algorithm that exploits quantum parallelism to
simulate randomness in a quantum system. In our scheme, all
possible realizations of the random parameters are encoded quantum
mechanically in a superposition state of an auxiliary system. We
show how our algorithm allows for the efficient simulation of
dynamics of quantum random spin chains with known numerical
methods. We also propose an experimental realization based on
atoms in optical lattices in which disorder could be simulated in
parallel and in a controlled way through the interaction with
another atomic species.
\end{abstract}

\date{\today}
 \maketitle

One of the most remarkable features of quantum mechanics is that
it allows for the creation of superposition states, making the
ambitious dream of performing many tasks at the same time a real
possibility. This extraordinary concession lies at the heart of
quantum computation and is the basis of all quantum algorithms
developed so far \cite{Bennett93}. In this letter we present a new
algorithm that exploits this quantum parallelism to simulate in
parallel many different evolutions of a quantum system. Our
motivation is the simulation of physical systems whose
understanding requires to study their behavior under many
different Hamiltonians. An important example of such systems are
quantum random systems (QRS) \cite{Fisher88}. For them certain
parameters of the Hamiltonian (e.g., interaction couplings,
potential strengths) are random (classical) variables. Therefore
the exact simulation of their dynamics requires to perform many
evolutions, one for each realization of the set of random
variables. QRS have captured a lot of attention in the last
decades \cite{Fisher88, Belitz94, Fisher95}.  The presence of
randomness can dramatically change the behavior of quantum
many-body systems, leading to fascinating phenomena
\cite{Fisher88}. Moreover, the answer to puzzles as the unusual
transport properties of high temperature superconductor materials
is inextricably tied to the understanding of phase transitions and
transport in the presence of disorder \cite{Fisher92}. On the
theoretical side, the understanding of QRS is hindered  by the
fact that the number of required simulations for an exact
calculation scales exponentially with the number of random
parameters \cite{Fisher04}. On the experimental side, one of the
big challenges is the creation of randomness in a controlled way.
Here, atomic systems in optical lattices \cite{Cirac04, Bloch04},
highly versatile and controllable, are one of the most promising
candidates.

In this letter we present an algorithm that allows to simulate
dynamics (and ground state properties) of a QRS within one single
time evolution in which the system is put into interaction with an
auxiliary system. The key idea is that all possible realizations
of the set of random classical parameters  are encoded quantum
mechanically in a superposition state of the auxiliary system.
Choosing the interaction with the ancilla in the appropriate way,
all possible quantum evolutions of the QRS  are simulated in
parallel. As a particular case, adiabatic evolution with the
ancilla can simulate at once all possible ground states of the
QRS. As one of the main results of this work, our algorithm
establishes an exact mapping between a QRS and a certain
interacting (non-random) system. This equivalence opens a new path
both to the numerical and experimental simulation of QRS. On the
numerical side, it allows for the efficient simulation of QRS
within the framework of numerical methods that simulate the
corresponding interacting systems efficiently. For example, for
the case of a quantum random spin chain  we will show that the
problem is mapped onto the simulation of the time evolution of a
one-dimensional (1D) lattice system. Recently, numerical methods
inspired in density matrix renormalization group (DMRG)
\cite{White92} and matrix-product-state (MPS) \cite{Ostlund95}
techniques have been developed to efficiently simulate the time
evolution of 1D lattice systems \cite{Vidal03, Verstraete04,
White-Daley04}. Here we will show how to implement the efficient
simulation of quantum random spin chains within the methods
introduced in \cite{Verstraete04}. On the experimental side our
scheme opens the possibility of simulating randomness in parallel
through the interaction with an auxiliary quantum system.
Moreover, conversely, it allows to interpret current interacting
experimental schemes as potential simulators of certain random
equivalent problems. We propose an experimental scheme in which a
variety of disordered phases as quantum glasses
\cite{Fisher89,Sanpera04} or Anderson insulator phases
\cite{Hofstetter05} could be simulated in current experiments with
optical lattices \cite{Bloch02, Esslinger05}.

 {\em The algorithm}. Let us consider a quantum system with
Hilbert space $\mathcal{H}$ that evolves accordingly to a
Hamiltonian $H(r_1, \ldots, r_n)$ where $r_1, \ldots, r_n$ are
random variables that take values within a finite discrete set,
$r_\ell \in \Gamma_{\ell}=\{\lambda_1^{\ell}, \ldots \
\lambda^{\ell}_{m_{\ell}}\}$, with a probability distribution
given by $p(r_1, \ldots, r_n)$. In order to simulate exactly the
dynamics of such a system one would need to perform
$\prod_{\ell=1}^n m_{\ell}$ simulations, one per each possible
realization of the set of random variables $\mathbf{r}={(r_1,
\ldots r_n)}$. For each realization the system evolves to a
different state $\vert \psi_{\mathbf{r}}(t)
\rangle=e^{-iH(\mathbf{r})t /\hbar} \vert \psi_0 \rangle$, where
$\vert \psi_0 \rangle$ is the initial state. Given this set of
evolved states and a physical observable $\hat{O}$, one is
typically interested in the average of the expectation values of
that observable in the different evolved states, that is, in
quantities of the form:
\begin{equation}
{\big \langle} \langle \hat{O}(t) \rangle {\big \rangle} :=
\sum_{\mathbf{r}} p(\mathbf{r}) \langle \psi_{\mathbf{r}}(t)\vert
\hat{O} \vert \psi_{\mathbf{r}}(t) \rangle.
\label{valores_esperados_dobles}
\end{equation}
On the following we describe an algorithm that allows to simulate
in parallel all possible time evolutions of the random system
described above. We consider an auxiliary system with Hilbert
space $\mathcal{H}_a$ and a Hamiltonian acting on
$\mathcal{H}\otimes \mathcal{H}_a$ of the form
$\widetilde{H}=H(\hat{R_1}, \ldots, \hat{R_n})$, where $\hat{R_1},
\ldots, \hat{R_n}$ are operators that act in $\mathcal{H}_a$,
commute with each other and have spectra $\Gamma_1, \ldots
,\Gamma_n$. Note that we have replaced the set of random variables
$\mathbf{r}$ by a set of quantum operators $\mathbf{\hat{R}}$ with
the same spectra. The algorithm works as follows. 1) {\em
Initialization}. Let us prepare the auxiliary system in an initial
superposition state of the form:
\begin{equation}
\vert \psi_a \rangle= \sum_{\mathbf{r}}
\sqrt{p(\mathbf{r})}\,\,\vert \mathbf{r}\rangle,
\label{estado_ancila}
\end{equation}
where the states $\vert \mathbf{r}\rangle$ are simultaneous
eigenstates of the set of operators $\mathbf{\hat{R}}$, with
$\hat{R_{\ell}}\vert \mathbf{r}\rangle=r_\ell \vert
\mathbf{r}\rangle$. Each state $\vert \mathbf{r} \rangle$ is
therefore in one to one correspondence with one realization of the
the set of random variables $\mathbf{r}$, its weight in the
superposition state (\ref{estado_ancila}) being equal to the
probability with which the corresponding realization occurs for
the random system. 2) {\em Evolution}. We evolve the initial state
of the composite system $\vert \psi_0 \rangle \otimes \vert \psi_a
\rangle$ under the Hamiltonian $\widetilde{H}$. The evolved state
is
\begin{equation}
\vert \Psi(t)\rangle = \sum_{\mathbf{r}}\!
\sqrt{p(\mathbf{r})}\,\vert \psi_{\mathbf{r}}(t) \rangle \otimes
\vert \mathbf{r}\rangle.
\label{estado_evolucionado_sistema_mas_ancila}
\end{equation}
This superposition state contains the complete set of evolved
states we are interested in. 3) {\em Read-out}. In order to obtain
the quantities (\ref{valores_esperados_dobles}) we just need to
measure the observable $\hat{O} \otimes 1$,
\begin{eqnarray}
\langle \Psi (t)\vert \hat{O} \otimes 1 \vert \Psi (t)
\rangle=\big{\langle} \langle \hat{O}(t) \rangle \big{\rangle}.
\end{eqnarray}
The algorithm above allows us, in particular, to obtain the
averaged properties of a random system over the collection of all
possible {\em ground states}. Let us assume that the interaction
between the system and the ancilla is introduced adiabatically, so
that the Hamiltonian is now $\widetilde{H}(t)= H
(\beta(t)\mathbf{\hat{R}})$, where $\beta (t)$ is a slowly varying
function of time with $\beta(0)=0$, $\beta(T)=1$, $T$ being the
time duration of the evolution. If the system is prepared in the
ground state of the Hamiltonian $H(\mathbf{0})$, the algorithm
above will simulate in parallel all possible adiabatic paths, so
that the composite superposition state
(\ref{estado_evolucionado_sistema_mas_ancila}) will contain all
possible ground states of the random system \cite{ground-states}.

Additionally, the scheme above can be easily extended for the
computation of other moments of the distribution of physical
observables (higher than (\ref{valores_esperados_dobles})), which
are sometimes important in the understanding of QRS
\cite{Fisher95}. For example, quantities like $\big{\langle}
\langle \hat{O}^2 \rangle- \langle \hat{O} \rangle ^2
\big{\rangle} $, can be computed by using an additional copy of
the system \cite{Paredes05}.

{\em Numerical implementation}. The algorithm described above
reduces the simulation of a quantum random system to the
simulation of an equivalent non-random interacting problem. This
exact mapping allows us to integrate the simulation of randomness
in quantum systems within the framework of numerical methods that
are able to efficiently simulate the corresponding interacting
problem.   As an illustrative example we consider the case of a 1D
spin $s=1/2$ system with random local magnetic field. The
Hamiltonian of the system is:
\begin{equation}
H(b_1, \ldots, b_N)=H_0+B\sum_{\ell=1}^N b_\ell S_\ell^{z},
\label{ham_campo_mag_random}
\end{equation}
where $H_0$ is a short range interaction Hamiltonian,
$\mathbf{b}=(b_1, \ldots, b_N)$ is a set of classical random
variables that take values $\{1/2,-1/2\}$ with probability
distribution $p(\mathbf{b})$. Following the algorithm above the
$2^N$ simulations required for the exact simulation of the
dynamics (or the ground-state properties) of this random problem
can be simulated in parallel as follows. We consider an auxiliary
1D spin $\sigma=1/2$ system. We prepare this ancilla in the
initial state $\vert \psi_a \rangle= \sum_{\mathbf{b}}
\alpha_{\mathbf{b}}\vert \mathbf{b}\rangle$, where the states
$\vert \mathbf{b}\rangle$ have all $z$ components of the $N$ spins
well defined, $\widehat{\sigma}^z_\ell \vert
\mathbf{b}\rangle=b_{\ell}\vert \mathbf{b}\rangle$, and
$\alpha_{\mathbf{b}}=\sqrt{p(\mathbf{b})}$. The entangled
properties of the state of the ancilla reflect the classical
correlations among the random variables. For example, for a
uniform distribution of the random field, $p(\mathbf{b})=1/2^N$,
the state of the ancilla is just a product state, $\vert \psi_a
\rangle \propto \left( \vert \uparrow \rangle + \vert \downarrow
\rangle \right)^{\otimes N}$. We evolve the system and the ancilla
under the interaction Hamiltonian
\begin{equation}
\widetilde{H}=H(\widehat{\sigma}^z_1, \ldots,
\widehat{\sigma}^z_n)=H_0+\beta \sum_\ell \widehat{\sigma}^z_\ell
\widehat{S}_\ell^{z}. \label{ham_dos_cadenas}
\end{equation}

Here, $\beta=B$ if we want to simulate dynamics under Hamiltonian
(\ref{ham_campo_mag_random}), and $\beta$ is a slowly varying
function of time with $\beta(0)=0$ and $\beta(T)=B$ for the
simulation of the ground state properties. We have then reduced
the simulation of the random problem to that of the time evolution
of two coupled spin $1/2$ chains with Hamiltonian
(\ref{ham_dos_cadenas}). This problem is equivalent to a 1D
lattice problem of $N$ sites with physical dimension $d=2 \times
2$, which can be easily incorporated to the framework of the
numerical methods introduced in \cite{Vidal03, Verstraete04}.
Implementation of the scheme above is as follows.

\begin{figure}[b]
\includegraphics[width=0.85\linewidth,height=12.6cm]{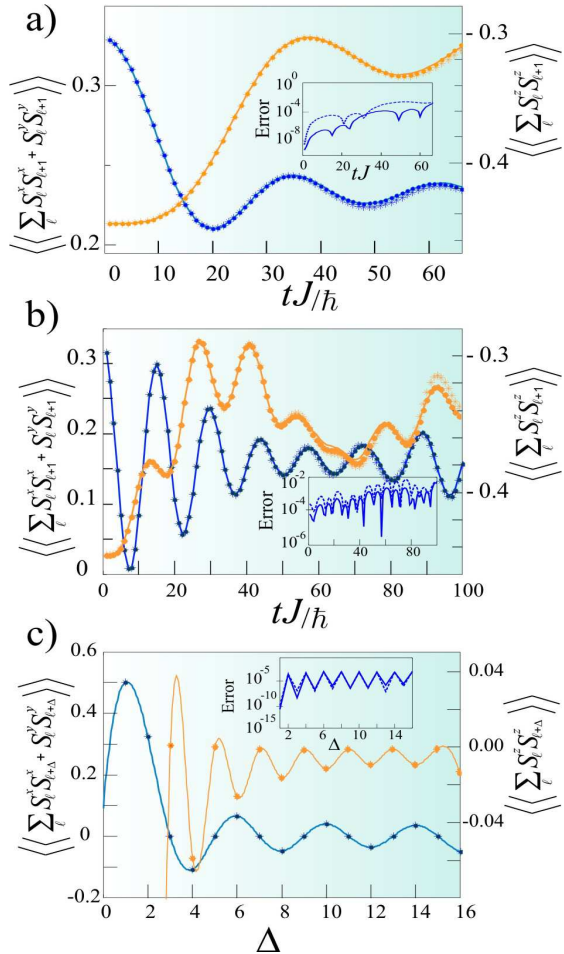}
\caption{Comparison of our numerical simulation of a random field
XY spin chain with an exact calculation. We plot the exact results
(full line) together with the numerical results using MPS with
$D=20$ (circles) and $D=30$ (stars). The relative errors are
plotted in the insets. Figures (a) and (b) show the time evolution
of the averaged correlations ${\big \langle} \langle \sum_\ell
S^x_{\ell} S^x_{\ell+1}+S^y_{\ell} S^y_{\ell+1} \rangle {\big
\rangle}$ (blue) and ${\big \langle} \langle \sum_\ell S^z_{\ell}
S^z_{\ell+1} \rangle {\big \rangle}$ (orange). The evolution
Hamiltonian is (\ref{ham_campo_mag_random}) with (a) $B_0=0$,
$J/B=-2$ and $p({\bf b})=1/2^N$, and (b) $B_0=0$, $J/B=-4$ and
$p(\bf{b})=\vert \alpha_{{\bf b}} \vert^2$, where $\alpha_{{\bf
b}}$ are given by $\vert \psi_a \rangle= \sum_{\mathbf{b}}
\alpha_{\mathbf{b}}\vert \mathbf{b}\rangle$, and $\vert \psi_a
\rangle$ is the ground state of $H_0$ with $B_0/J=1.4$. For both
figures the chain is prepared initially in the ground state of
$H_0$ with $B_0=0$. The time step for the numerical simulations is
$\Delta t=0.01J/\hbar$. Fig. (c) shows the correlation functions
${\big \langle} \langle \sum_\ell S^x_{\ell}
S^x_{\ell+\Delta}+S^y_{\ell} S^y_{\ell+\Delta} \rangle {\big
\rangle}$ (blue) and ${\big \langle} \langle \sum_\ell S^z_{\ell}
S^z_{\ell+ \Delta} \rangle {\big \rangle}$ (orange) averaged over
all possible ground states of Hamiltonian
(\ref{ham_campo_mag_random}) for $B_0=0$, $J/B=-1$ and $p({\bf
b})=1/2^N$ as a function of the distance $\Delta$ between spins.
The numerical simulation performs an adiabatic evolution with
Hamiltonian (\ref{ham_dos_cadenas}) with $\beta (t)=Bt/T$ and
$T=100J/\hbar$.}
\end{figure}

\begin{figure}[b]
\begin{center}
\includegraphics[width=\linewidth]{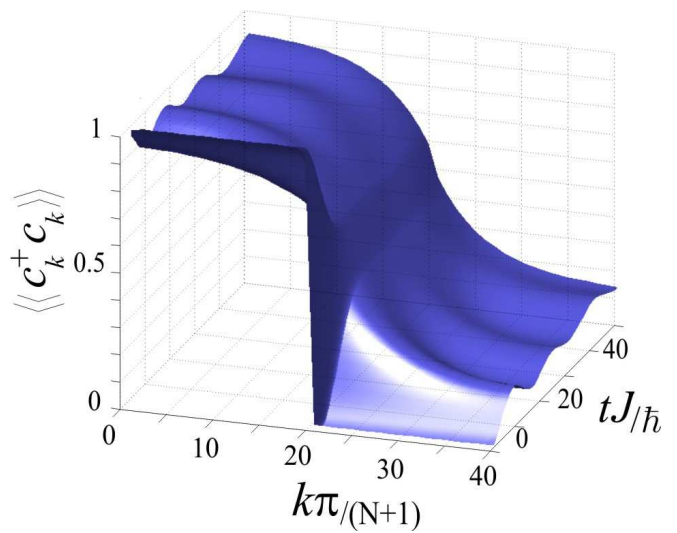}
\caption{Numerical simulation of the time evolution of a random
field XY spin chain with $N=40$. We show the correlation function
${\big \langle} \langle c^{\dagger}_k c_k \rangle {\big \rangle}$
as a function of time and momentum $k$. Here $c_k \propto
\sum_\ell sin(k \ell) \widetilde{c}_\ell$, $k=\frac{\pi}{N+1},
\ldots, \frac{\pi N} {N+1}$ and $\widetilde{c}_\ell=\prod_{\ell <
\ell'} S^z_{\ell'} (S^x_{\ell} + i S^y_{\ell})$ are the fermionic
operators given by the Jordan-Wigner transformation
\cite{Sachdev}. The evolution Hamiltonian is
(\ref{ham_campo_mag_random}) with $H_0$ being the XY Hamiltonian
with $B_0=0$, $B/J=-4$ and $p({\bf b})=1/2^N$. The initial state
$\vert \psi_0 \rangle$ is the ground state of $H_0$. As the system
evolves in time the initially sharp Fermi sea disappears.}
\end{center}
\end{figure}

\begin{figure}[h]
\begin{center}
\includegraphics[width=\linewidth]{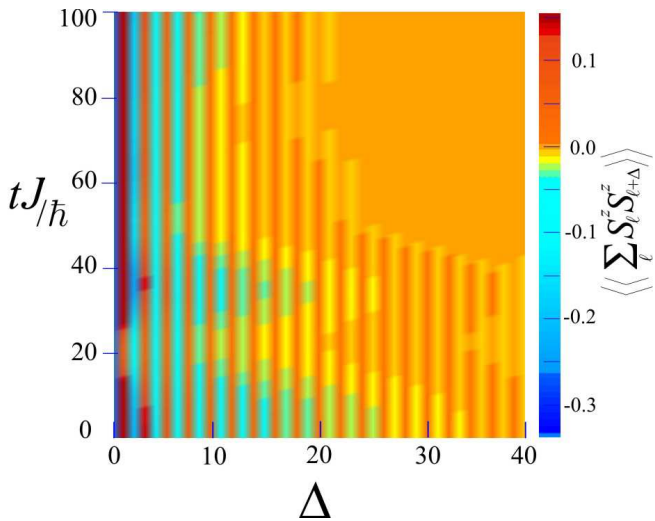}
\caption{Numerical simulation of the time evolution of a random
field Heisenberg spin chain with $N=40$. We show the correlation
function ${\big \langle} \langle \sum_\ell S^z_{\ell} S^z_{\ell+
\Delta} \rangle {\big \rangle}$ as a function of time and the
separation $\Delta$ between the spins. The evolution Hamiltonian
is (\ref{ham_campo_mag_random}) with $H_0$ being the
antiferromagnetic Heisenberg Hamiltonian, $B/J=-4$ and $p({\bf
b})=1/2^N$. The initial state $\vert \psi_0 \rangle$ is the ground
state of $H_0$. As the system evolves in time the
antiferromagnetic correlations are smeared out.}
\end{center}
\end{figure}
1) Let the state
$\vert \Psi(0) \rangle$ be the MPS with dimension $D$ that better
approximates the initial state $\vert \psi_0 \rangle \otimes \vert
\psi_a \rangle$,
$\vert \Psi(0) \rangle=\sum_{s_1, \ldots, s_N=1}^d \mathrm{Tr}
\left( A_1^{s_1}\ldots A_N^{s_N} \right) \vert  s_1 \ldots s_N
\rangle$.
Here, the $A$'s are matrices whose dimension is bounded by $D$ and
$d=4$. 2) We evolve $\vert \Psi(0) \rangle$ under Hamiltonian
(\ref{ham_dos_cadenas}). As in \cite{Verstraete04} we take a small
time step $\Delta t$ and compute $\vert \Psi(\Delta t) \rangle$
exactly, a state for which the dimension of the matrices $A$ will
be typically larger. Following \cite{Verstraete04} we then
``truncate'' the matrices in an optimal way and use the
``truncated'' state to compute the next time step. 3) At any time
$t$ we can efficiently determine the quantities $\langle \langle
O_1 \ldots O_N \rangle \rangle$ as $\langle O_1 \otimes 1 \ldots
O_N \otimes 1 \rangle$, which can be efficiently computed for MPS
\cite{Verstraete04}.

In order to test the efficiency of the simulation scheme above we
have compared it with an exact calculation for the case in which
$H_0$ is an XY model Hamiltonian, $H_0=-J\sum_{\ell=1}^N  S_\ell
^x S_{\ell+1} ^x +S_\ell ^y S_{\ell +1} ^y + B_0 \sum_{\ell=1}^N
S_\ell ^z$. The exact averages (\ref{valores_esperados_dobles})
are calculated in the following way. For each realization of the
magnetic field the evolved state (or ground state) of the
Hamiltonian (\ref{ham_campo_mag_random}) is computed exactly using
fermionization techniques \cite{Sachdev, Paredes04}. The
quantities (\ref{valores_esperados_dobles}) are then determined by
averaging over the expectation values of the $2^N$ states
obtained. The comparison is shown in Fig. 1 for $N=16$ and
different correlations functions averaged over the collection of
evolved states (Fig. 1(a) and 1(b)) and ground states (Fig. 1(c)),
both for a uniform probability distribution (Fig. 1(a)) and a
correlated one (Fig. 1(b)). Using MPS with $D=20$ we obtain a very
good accuracy (see insets in Fig. 1). As in \cite{Vidal03,
Verstraete04} we have two sources of error. One is the Trotter
expansion. Here the error can be decreased by considering smaller
time steps. The other source of error is the truncation of the MPS
to a smaller dimension. For the simulations we have performed this
error did not grow much with the size of the system. Using the
numerical scheme above we have performed simulations of the
dynamics of random field XY (Fig. 2) and Heisenberg (Fig. 3)
chains with $N=40$.

The scheme above can be easily extended for the numerical
simulation of spin chains with {\em random couplings}. For
example, for the case of a Heisenberg chain with Hamiltonian
$H(J_1, \ldots, J_N)=\sum_\ell J_\ell \mathbf{S}_\ell \cdot
\mathbf{S}_{\ell+1}$, the problem is mapped to the simulation of
the time evolution of two spin chains under the three-body
interaction Hamiltonian $\widetilde{H}=H(\widehat{\sigma}^z_1,
\ldots, \widehat{\sigma}^z_n)= \sum_\ell \widehat{\sigma}^z_\ell
\mathbf{S}_\ell \cdot \mathbf{S}_{\ell+1}$.

{\em Experimental proposal}. Using the ideas of the algorithm
above we present an experimental scheme for atoms in optical
lattices that could be use as a simulation protocol for a variety
of disordered phases. We consider a system of atoms $b$ (bosons or
fermions) in an optical lattice (3D, 2D, or 1D) in a certain state
$\vert \psi_0 \rangle$. We consider another system of atoms $a$
(e.g, another spin state or atomic species) which experiences an
independent lattice potential \cite{Mandel03}. We consider a
situation in which the lattice potentials of atoms $a$ and $b$ are
initially shifted in such a way  that there is no interaction
between the two systems. We proceed as follows: 1) We prepare
atoms $a$ is a certain state $\vert \psi_a \rangle$. This state
can always be written in a Fock basis as $\vert \psi_a
\rangle=\sum_{n_1 \ldots n_M} \alpha_{n_1, \ldots n_M} \vert n_1
\ldots n_M \rangle$, where $n_1, \ldots n_M$ are the occupation
numbers of the $M$ lattice sites and the $\alpha$'s are certain
complex coefficients. We then suddenly ramp up the lattice for
atoms $a$ (so that tunneling processes are instantaneously
suppresed), and shift it so that the interaction with atoms b is
instantaneously switched on. 2) We let the composite system
evolve. The Hamiltonian that governs the evolution is
\begin{equation}
\widetilde{H}=H_b+U_{ab}\sum_{\ell=1} ^M \hat{n}^a_\ell
\hat{n}^b_\ell,
\end{equation}
where $H_b$ is the Hubbard Hamiltonian for atoms $b$,
$\hat{n}^a_\ell, \, \hat{n}^b_\ell$ are the local density
operators for atoms $a$ and $b$, and $U_{ab}$ is the interaction
coupling between atoms $a$ and $b$. 3) We finally measure the
system $b$. According to the algorithm developed above, this
interacting experimental scheme is simulating in parallel all
possible dynamics of atoms $b$ under the random Hamiltonian:
\begin{equation}
H(V_1, \ldots , V_M)=H_b+U_{ab}\sum_{\ell=1} ^M V_\ell
\hat{n}^b_\ell,
\end{equation}
\begin{figure}[b]
\begin{center}
\includegraphics[width=\linewidth]{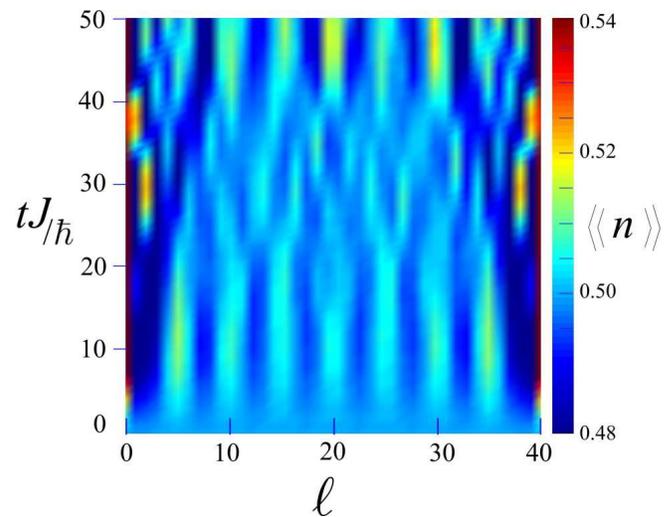}
\caption{Numerical simulation of the time evolution of the
averaged density of a Tonks gas in an optical lattice with $M=40$
sites and $\nu=1/2$ in the random potential generated by another
Tonks gas with $\nu=1/5$.}
\end{center}
\end{figure}
where $V_1, \ldots, V_M$ are random potential strengths that take
values within $\{0,1, \ldots, N_a \}$ with a probability
distribution given by $p(V_1, \ldots, V_M)=\vert \alpha_{V_1,
\ldots V_N}\vert ^2$, $N_a$ being the number of $a$ atoms.
Choosing the initial state of atoms $a$ in the appropriate way we
can tune the probability distribution of the random potential for
atoms $b$. Changing the entanglement properties of the state in
which atoms $a$ are prepared we will change the correlation
properties among the random local potentials. As well, the
intensity of the disorder potential can be tuned by varying the
interaction strength $U_{ab}$ (e.g., shifting the lattices). This
allows to simulate a large variety of disordered phases. As
opposite to the classical simulation of randomness (e.g.,with
speckle lasers \cite{Horak98}) our quantum mechanical approach
allows to simulate all possible evolutions of the random system in
one single run of the experiment. Concerning measurements, note
that in an experiment we will typically have many copies of the
system (an array of 2D or 1D systems) so that the outcomes will be
already averaged over all copies. As an example of current
interest let us consider the case in which atoms $a$ and $b$ are
initially prepared in two independent Tonks states
\cite{Paredes04} with filling factor $\nu_a$ and $\nu_b$. For this
case the experimental scheme above would simulate the dynamics of
a Tonks gas in the presence of the random potential generated by
another Tonks gas. Using the numerical scheme above we have
simulated this situation for $M=40$ sites and $\nu_b=1/2$,
$\nu_a=1/5$. Interestingly, the averaged density of atoms $b$
shows localization of atoms $b$ in the regions in which atoms $a$
are most probably absent (see Fig. 4).

In conclusion, we have presented an algorithm that allows to
simulate (classical) randomness in quantum many body systems via a
single quantum mechanical problem in which quantum interactions
allow all possible quantum paths of the random system to occur
simultaneously. Our scheme opens new possibilities in the
numerical and experimental simulation of QRS.

We thank M. A. Mart\'{\i}n-Delgado for helpful discussions. Work
supported in part by DFG, EU projects and Bayerische
Staatsregierung.


\begin{thebibliography}{99}
\bibitem{ground-states}As long as there are no level crossings.
\bibitem{Paredes05}
Let us consider two independent copies of the system both
interacting with the ancilla via the Hamiltonian
$\widetilde{H}^{(1)}+ \widetilde{H}^{(2)}$. The 3-party system
evolves then to the state $\vert \Psi (t)
\rangle=\sum_{\mathbf{r}}\! \sqrt{p(\mathbf{r})}\,\vert
\psi_{\mathbf{r}}(t) \rangle^{(1)} \otimes \vert
\psi_{\mathbf{r}}(t) \rangle^{(2)}\otimes \vert
\mathbf{r}\rangle.$ We then measure $\langle \Psi (t)\vert
\hat{O}^{(1)} \otimes \hat{O}^{(2)}\otimes 1 \vert \Psi (t)
\rangle=\big{\langle} \langle \hat{O} \rangle^2 \big{\rangle}.$
\end{thebibliography}
\end{document}